\documentclass[journal]{IEEEtran}
\IEEEoverridecommandlockouts
\usepackage{cite}
\usepackage{amsmath,amssymb,amsfonts}
\usepackage{algorithmic}
\usepackage{graphicx}
\usepackage{textcomp}
\usepackage{xcolor}
\usepackage{color, colortbl} 
\definecolor{DarkMaroon}{rgb}{.61,.2,.4}
\definecolor{MedDarkMaroon}{rgb}{.88,.64,.76}
\definecolor{MedLightMaroon}{rgb}{.92,.77,.85}
\definecolor{LightMaroon}{rgb}{.96,.89,.93}
\usepackage{soul}

\usepackage{todonotes}
\usepackage{float}
\def\BibTeX{{\rm B\kern-.05em{\sc i\kern-.025em b}\kern-.08em
    T\kern-.1667em\lower.7ex\hbox{E}\kern-.125emX}}
\begin{document}

\title{Test and Evaluation Framework for Multi-Agent Systems of Autonomous Intelligent Agents \\
\thanks{This paper includes funded research conducted through the System Engineering Research Center.}
}

\author{
 \IEEEauthorblockN{
    Erin Lanus\IEEEauthorrefmark{1}, 
	Ivan Hernandez\IEEEauthorrefmark{1}, 
    Adam Dachowicz\IEEEauthorrefmark{2}, 
    Laura Freeman\IEEEauthorrefmark{1},  
    Melanie Grande\IEEEauthorrefmark{2}, 
    Andrew Lang\IEEEauthorrefmark{2},
    Jitesh H. Panchal\IEEEauthorrefmark{2},
    Anthony Patrick\IEEEauthorrefmark{3}, 
    Scott Welch\IEEEauthorrefmark{1} 
  }	

  \IEEEauthorblockA{
    \IEEEauthorrefmark{1}Virginia Tech, Arlington, VA 22309, USA
  }
  \IEEEauthorblockA{
    \IEEEauthorrefmark{2}Purdue University, West Lafayette, IN 47907, USA
  }
    \IEEEauthorblockA{
    \IEEEauthorrefmark{3}George Mason University, Fairfax, VA 22030, USA
  }
}

\maketitle
\begin{abstract}
Test and evaluation is a necessary process for ensuring that engineered systems perform as intended under a variety of conditions, both expected and unexpected. In this work, we consider the unique challenges of developing a unifying test and evaluation framework for complex ensembles of cyber-physical systems with embedded artificial intelligence. We propose a framework that incorporates test and evaluation throughout not only the development life cycle, but continues into operation as the system learns and adapts in a noisy, changing, and  contended environment. The framework accounts for the challenges of testing the integration of diverse systems at various hierarchical scales of composition while respecting that testing time and resources are limited. A generic use case is provided for illustrative purposes and research directions emerging as a result of exploring the use case via the framework are suggested. 
\end{abstract}


\section{Introduction}
 The United States engages in numerous strategic initiatives to increase the use of Artificial Intelligence (AI) to support strategic priorities. Achieving complex mission needs requires AI to be integrated and deployed in \textit{multi-agent systems of autonomous intelligent agents (AIAs)}. These systems, if proven to be reliable, trustworthy, and safe, have the potential to be used in high-stakes contexts, often with lack of human intervention and under changing mission and environmental needs. This research is motivated by the challenge of testing AIAs as compared to static, deterministic systems.
 
 Test and evaluation (T\&E) of multi-agent systems of AIAs presents unique challenges due to the dynamic environments of the agents, adaptive learning behaviors of individual agents, the complex interactions among the agents, the complex interactions between agents and the operational environment, the difficulty in testing black-box machine learning models, and rapidly evolving AI algorithms. Currently, no unifying framework exists for T\&E of multi-agent systems of AIAs. Existing frameworks for T\&E of complex engineered systems \cite{DAG} fail to account for these unique challenges.

T\&E is a difficult topic to study as different fields of engineering have evolved their test strategies to meet the specific needs of that field.  For example, the reliability community has a robust literature on reliability testing \cite{Reliabilityoverview}, the software community has methods for software testing \cite{ACTS}, and manufacturing has methods for testing the consistence of their processes \cite{lowry1995review}.  However, complex systems require the integration of many methods to fully characterize system capabilities and understand how they will perform in the actual operational environment.  The United States Department of Defense (DoD) has a mature T\&E process due to the nature of the technologies they must ensure perform adequately and are safe before fielding.  These test processes are documented in the Defense Acquisition Guidebook \cite{DAG} and provide a comprehensive overview of these processes, but notably missing is any guidance on how processes should account for AIA challenges. 

 The development of multi-agent systems of AIAs involves taking an interdisciplinary approach, with each discipline providing its own methods, tools, techniques, priorities, and expertise. Consequently, the collection of accompanying T\&E strategies across a multi-agent system is heterogeneous, and it is not clear how individual T\&E strategies for components or subsystems should be combined to provide a comprehensive T\&E framework for a multi-agent system of AIAs.  Furthermore, new system capabilities, applications, properties, and behaviors emerge at the intersection of \textit{multi-agent}, \textit{autonomous}, and \textit{intelligent} systems.  
 
 New constructs within T\&E are necessary to facilitate addressing these new challenges in a manageable framework used for systematic analysis.  For example, a multi-agent system-of-systems requires additional testing across a hierarchical scale as component subsystems are integrated. Testing must be conducted on ``local'' factor levels specific to an individual agent as well as ``global'' factor levels representing the combined interactions of the different agents along with environmental conditions. The dynamic nature of multi-agent interactions can have effects apparent at the lower hierarchical scale of a given agent and also produce emergent phenomena at higher hierarchical scale. Additionally, an increase in the number of agents, each with its own parameters, exponentially increases the number of tests that might be conducted to support a comprehensive evaluation. Finally, these AIAs will have the ability to learn over time, so test strategies that continuously evaluate both the local and global scale over time are needed. The increase in agents and parameters requires more time and resources for conducting T\&E. We hypothesize that improving efficiency and coverage in a distributed, dynamic learning environment is essential to a T\&E framework for multi-agent systems of AIAs.

 In this work, we propose a unifying framework for T\&E of systems of multiple AIAs to guide the systematic development of test plans. The framework is informed by three major concepts that address the unique needs of this context: 1) field of study, 2) hierarchy of test, and 3) test plan efficiency. Collectively and along with an expanded systems engineering verification and validation model, these concepts describe how to define a \emph{slice} of the process during a phase of the system design, development, and deployment life cycle in order to identify goals of the test and inform creation of a comprehensive test plan. The rest of the paper is organized as follows. An illustrative use case of a satellite system is presented in \S~\ref{sec:use_case}. The VTP model on which the framework is built is presented in \S~\ref{VTP}. Testing procedures drawn from fields of study utilized in building these systems of systems and how they are addressed in the framework are discussed in \S~\ref{field}. How to conduct integration testing as components are merged into subsystems and subsystems into systems as a hierarchical approach to testing is considered in \S~\ref{hierarchy}. Maximizing knowledge gained with limited testing resources as the goal of test plan efficiency is discussed in \S~\ref{efficiency}.  The complete framework is given in \S~\ref{framework}. Finally, conclusions and research directions emerging as a result of exploring the use case via the framework are suggested \S~\ref{conclusions}.

\section{Illustrative Use Case}\label{sec:use_case}

\begin{figure}[htb]
    \centering
    \includegraphics[width=\columnwidth]{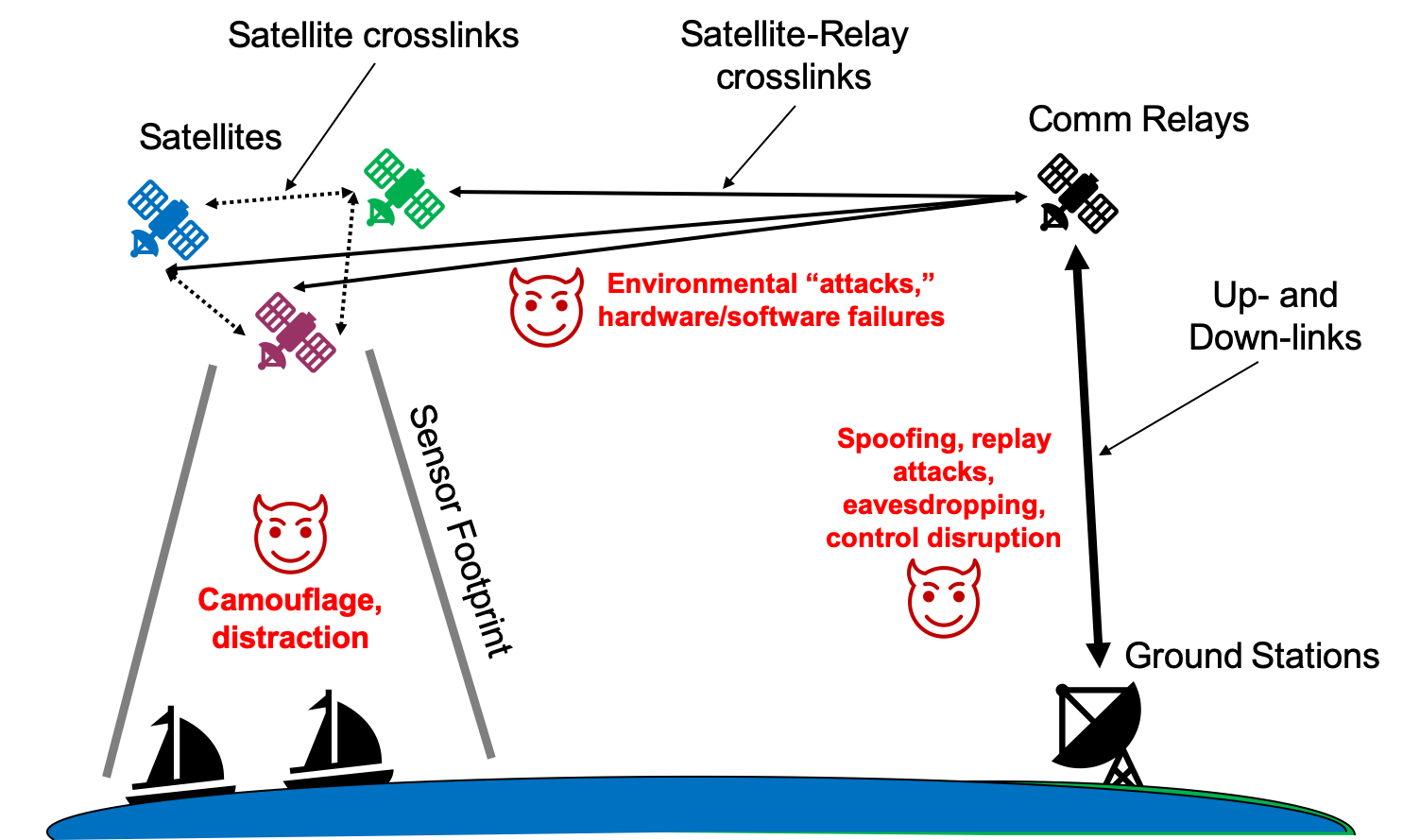}
    \caption{Illustrative use case of a multi-agent satellite system}
    \label{fig:usecase}
\end{figure}

 To provide an illustrative practical backdrop, we employ a generic use case of a satellite network composed of a heterogeneous set of AIAs reporting to a central controller and acting autonomously to conduct broad area search and point detection (see Fig.~\ref{fig:usecase}). At the local hierarchical scale, each satellite is composed of component subsystems such as sensors, actuators, and software including deterministic control software as well as AI software that is expected to change after deployment as a result of adapting to changing environments and knowledge acquisition. Each of these subsystems could be further decomposed into smaller components, such as a piece of hardware or a function within a program. At the global hierarchical scale, the system can be described by the types, number, and positions of each satellite, additional state information such as if the satellite has been damaged or its software has been compromised, and connectivity of each satellite with each other and the ground station. Last, operational environmental conditions can be specified such as lack of visibility, presence of adversarial powers and their capabilities, and presence and location of observational targets. 

\section{The VTP Model}\label{VTP}

\begin{figure*}
    \centering
    \includegraphics[width=\linewidth]{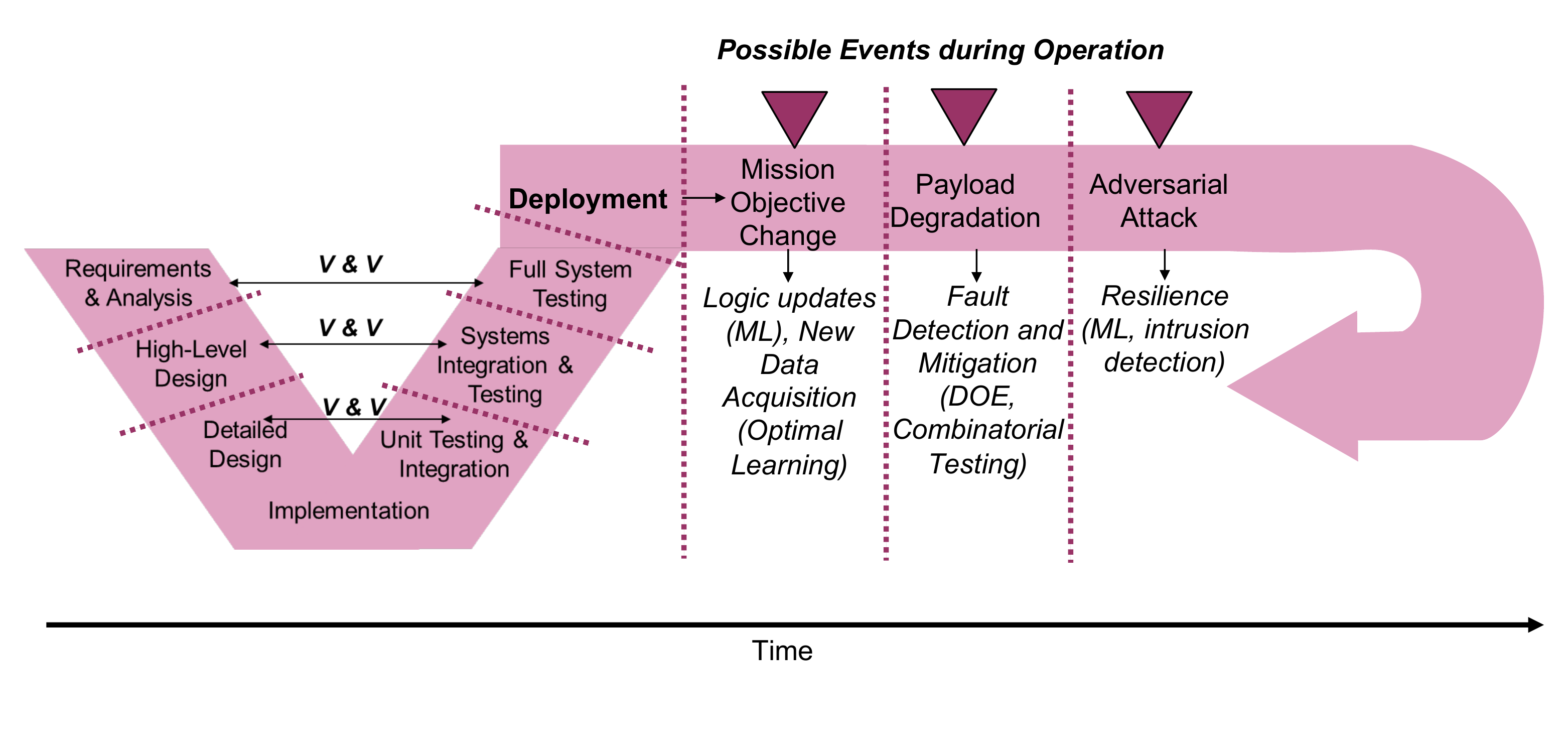}
    \caption{The VTP framework extends the ``Vee'' model to include testing throughout system deployment and a feedback loop}
    \label{fig:vtpmodel}
\end{figure*}

T\&E must be integrated throughout not only the system development process, but also the system life cycle. The Systems Engineering ``Vee'' model~\cite{buede2009} is a mature model of systems engineering that serves as a sound starting point for development of a unifying framework (see the left third of Fig.~\ref{fig:vtpmodel}). In the ``Vee'' model, each system-level in the hierarchy is paired with a corresponding level of verification and validation. Requirements and test plans are created from the beginning of the development process, rather than waiting until the entire system is developed. For example, at the beginning of the project when requirements for the entire system are identified, tests that verify system performance are designed though they are not executed until near the end. The system is thus designed top down, walking down the left side of the ``Vee'' and tests in the corresponding slice are defined simultaneously; however, test execution is conducted ``bottom up'' as the subsystems are built and the system is integrated, walking up the right side of the ``Vee.'' 

Since the ``Vee" model is based on the principle of hierarchical decomposition, certain assumptions of the ``Vee'' model do not hold for testing multi-agent system of AIAs. Specifically, most systems development life cycles assume well-defined phases, such as concept studies, technology development, preliminary design, final design, fabrication, assembly, test, launch, operations \& sustainment, and closeout~\cite{NASA_SE_Handbook}. The phases in the development process are akin to the phases in the the Software Engineering Waterfall Model that include requirement identification, design, implementation, testing, and maintenance. These models assume that all requirements against which the system will be tested are able to be listed during a requirements phase, and that requirements can be decomposed and traced to individual components or subsystems. Last, maintenance allows for ``fixing'' a product during deployment, but does not consider that behavior of the agents could change after deployment in the field. 

These assumptions do not hold for multi-agent systems of AIAs. Specifically, it may not be possible to define how the system should respond in all environmental conditions as the environment is constantly changing. For example, it may not be possible to define in advance the threat capabilities of advanced adversaries. Requirements may be achieved by multiple combinations of subsystems. A given task may be achieved by different ensembles of satellites given their heterogeneous capabilities and positions. After deployment, the satellite software may be modified by code pushes from the ground station, but it may also change through learning behaviors as the embedded AI acquires knowledge from interacting with its environment and through collaborative decision making with other satellites in the constellation.

Despite the challenges of sequential design or ``big design up front'' models, incremental or iterative models have limited applicability for systems that cannot be easily recalled once deployed. That is, development of subsystems such as components on an individual satellite or software may be incrementally designed, developed, and tested while on the ground, but a clear demarcation for deployment of the system occurs when the satellite is launched. Thus, our framework assumes the Systems Engineering ``Vee'' model up to deployment, though several iterations of the ``Vee'' could occur before the deployment cutoff. We then extend this model with a ``T'' phase to include testing throughout operation to detect or respond to events. Such events could include a change to the mission objective requiring a code push from the ground station to particular satellites and executing all pertinent tests from the previous phase as well as new tests to address the code changes. Embedded AI software may also adapt as a result of learning, and tests must be conducted to ensure the system is learning the ``right'' actions. As the system encounters debris or exhibits hardware degradation over time, tests must be run periodically to identify faults and enact mitigation strategies. Last, communication systems are inherently susceptible to adversarial attack, and software may include intrusion detection algorithms that may need to be updated with new signatures. Embedded AI should be tested for resilience to newly discovered vulnerabilities or maturity issues such as data drift.

Last, the model should process feedback. The ``P'' phase includes a loop back to the deployment phase due to changes in the system from learning for systems that are currently deployed. This loop can also extend further to inform the next phase of system development. The full VTP model on which the framework is built is presented in  Fig.~\ref{fig:vtpmodel}. In the figure, the dashed lines delineate the sub-phases within design, development, and deployment, and these sub-phases provide timeline context for a slice of the process.

\section{Field of study}\label{field} Each AIA is a cyber-physical system with embedded AI. T\&E of the composed system should draw from scientifically-based testing techniques for each subsystem and thus consider the peculiarities of testing for AI, deterministic software, electronic hardware, and mechanical systems \cite{lanotte2019formal, guo2019cyber}. 

For embedded AI, T\&E methods should  measure inherent weaknesses of the algorithms in use.  For example, the use of neural networks in learning requires an evaluation strategy that measures the performance sensitivity to transformations or noise added to the input. This is needed to detect attacks such as data poisoning and measure the impact on mission success in contended environments. 

The AI functionality of the software is also supported by other deterministic software, such as functions to receive input from sensors and control actuators in order to interact with the agent's environment, as well as to communicate with other agents within the multi-agent system. For code under development, white box methods that emphasize structural code coverage (e.g., statement, decision, condition, branch, and path coverage) can be employed. For commercial-out-of-the-box software (COTS) or vendor-supplied software, black-box techniques are required and include equivalence partitioning, boundary value analysis, decision tables, state transition testing, use case testing, and combinatorial techniques. 

In testing physical components, statistical analysis of response variables ascribe variance to different independent variables, or \emph{factors}, and to estimate the effect of different factor \emph{levels} on system performance. Design of experiments (DOE) is a systematic approach to choosing a set of test cases to ensure adequate coverage of the operational space, determine how much testing is enough, and provide an analytical basis for assessing test adequacy. DOE has also proven useful in testing complex systems with embedded software \cite{freeman2018informing}. Alternatively, optimal learning \cite{Powell_OptimalLearning} is an approach that begins with an initial set of tests to establish some information about the system. A Bayesian surrogate of the objective function is trained and the next tests are chosen based on a heuristic that combines exploration of the test space exhibiting the most uncertainty with exploitation of areas of the test space that maximize the objective function.

 Additionally, test strategies must evaluate the fully integrated system-of-systems and its ability to execute tasks autonomously. While each subsystem may reside within one field of study (e.g., software) and thus testing for the subsystem may follow known testing strategies for that field, the composed system spans multiple disciplines and T\&E  must consider all together. Some tests may need to be designed that account for the interaction of disparate systems. For example, a satellite may need to learn that visibility issues affecting a sensor can be overcome by changes to its position and thus tests requiring orchestrated interaction of sensors, embedded AI, control software, and actuators are all required to evaluate this behavior.
 
 Last, the above list is not exhaustive. Depending on the use case in which the AIAs are employed, additional fields may be considered. Testing AIAs teamed with humans or with significant human-in-the-loop components should consult the psychology testing literature for designing tests that address the variety of unique challenges such as attention issues along the human-computer interface and how humans and computers can express and understand collaborative goals. Further, even without human involvement, psychologist SME consultation can be beneficial to establish benchmarks for learning behaviors of AIAs and in testing for collaborative behaviors of ensembled AIAs.
 
\section{Hierarchy of test}\label{hierarchy} Rather than waiting until the complete system is built to test, testing is conducted throughout the development process in order to detect and correct flaws as early as possible. Unit testing is conducted on the smallest testable components, using simulated inputs or digital environments when necessary. As components are integrated into subsystems, the expectation is that components work as intended, but there may be interactions among them causing failures or interaction effects on performance. As an example, suppose some COTS control software is employed that does not expect the range of inputs produced by a given sensor on the satellite. When used in conjunction, the code may crash or unexpected behavior may occur. 

Combinatorial interaction testing (CIT) creates test suites to systematically detect failures caused by combinations of interacting components  up to a given size of interaction\cite{kuhn2015Combinatorial}. To use available tools such as the Automated Combinatorial Testing for Software to generate test suites \cite{ACTS}, testers must identify the components (synonymous with factors in DOE), the levels at which the components should be tested, the maximum interaction size called the \emph{strength}, and any \emph{constraints}, combinations of component levels that must not be tested together. This process requires that components or factors of interest are known to the tester, and continuous levels must be discretized in order to use CIT.

As a complex system of systems, a system that undergoes integration testing in one phase of testing becomes a component in the next phase. For example, at the local scale of the hierarchy, payload sensors, actuators, AI algorithm, and control software are component subsystems integrated into an AIA satellite, and integration verification and validation testing is performed to ensure the system performs as expected. Once deployed into a constellation, testing moves up the hierarchy, and each satellite becomes a subsystem within the global system. Interaction testing considers failures at the top system or mission scale, such as whether communication relays between satellites and the ground station are operational or whether the combined sensor footprint of the constellation is sufficient for a given tactical operation. Factors at the global scale of the hierarchy may also include environmental factors that cannot be controlled but can be observed during a test to evaluate their impact on mission performance, such as storms affecting visibility. Other factors may be simulated, such as adversarial attacks.

After deployment, failures occurring at the global scale of the hierarchy may be caused by interactions of subsystems immediately lower in the hierarchy along with global scale factors, such as storms affecting sensor visibility of a particular subset of satellites or interfering with communication with a ground station. CIT methods include fault localization techniques that can be used to identify interactions causing the fault \cite{kuhn2015Combinatorial}. In some cases, a problem with a component system further down the hierarchy may be causing an error to propagate up to higher systems. Techniques may need to be utilized to step down through the involved interactions at each scale of the hierarchy to locate the component causing the fault.

\section{Test plan efficiency}\label{efficiency} 
Rigorous testing under a variety of conditions provides a degree of assurance that the system will perform as expected. The test input space is defined by identification of system and environmental factors of interest and choosing a range of levels for each factor. The points chosen for a test plan thus result in some coverage of the multi-dimensional test input space. Each test incurs some cost and, in most scenarios, both time and resources for testing are limited. Different techniques prioritize efficiency versus coverage. 

A full factorial design from DOE is a test suite including all combinations of factors and levels, providing exhaustive testing and the ability to conduct an analysis of variance and characterise the effect of factor levels on system performance. In most complex systems with many factors with multiple levels, exhaustive testing is infeasible. Fractional factorial designs select a fraction of the factorial design with the result that some effects are aliased with others and variance cannot be fully attributed. The choice of fraction can lead to aliasing between main effects and higher order interaction effects that are not expected to be significant and thus provide sufficient system knowledge with fewer test points.

Optimal learning is a technique that also has the goal of reducing uncertainty and determining the best factor levels for improving system performance. By choosing tests adaptively via the exploration-exploitation heuristic policy, optimal learning can discover the ideal factor settings with fewer tests.  However, it relies on other knowledge such as physical laws and prior experience to ensure that the results from fewer tests can be used to make claims about performance in the rest of the operating space.

An often employed technique in CIT is to design a test suite from a covering array, a combinatorial array where the columns represent factors, the rows represent tests, and the values in each cell represent the level set for that factor in that test. Every combination of values for up to a given strength of interacting features appears in some test in the array; thus, a covering array guarantees to detect failures due to interactions of up to the strength specified. One row covers many interactions when the strength is smaller than the total number of factors and so produces a test suite with a fraction of the number of tests in the full factorial. The number of tests required grows logarithmically in the number of factors. While using a covering array as a test suite provides coverage of the test space in terms of interactions, it does not guarantee that the cause of the fault detected can be identified. 

Humans frequently interact with complex systems as part of the environment or system. Between-subject and within-subject experimental design strategies allow for the humans contribution to the test outcome to be characterized and separated in order to reduce aliasing with factors and levels in the environment \cite{charness2012experimental}.  Between subject designs randomly assign humans to unique combinations of factors and levels, while within-subject designs provide a basis for comparison by matching individuals to combinations of factors and levels.  

Each of the above strategies has unique strengths that are leveraged by our framework to systematically cover the input space at both local and global scale, integrate information across test slices, and determine the right test size at various stages of development and deployment. Specifically, finding flaws early in development can prevent extensive testing later that must step down to components lower in the hierarchy. Additionally, at the lower hierarchical scale, such as within a single function of code of a sensor, the test input space is smaller. Tests may be faster and more automatable, such as by running a script or using simulated inputs, and test oracles may be computable. As subsystems are composed, testing emphasis shifts to identifying faulty interactions among components. At the global scale, the fully composed system is tested in terms of achieving mission objective, and it may be necessary to focus on only the most critical factor levels.

These facts support conducting exhaustive testing at the component scale but sparser testing at the global scale. DOE full factorial or fractional factorial designs are likely best used for components lower in the hierarchy. CIT techniques can prevent exponential growth in test points as interactions among components are considered during integration. At the global scale, correct system behavior may not even be definable, particularly after deployment in changing conditions and after acquisition of knowledge by embedded AI components. In this case, optimal learning may be employed to estimate system boundaries and adaptively select tests as needed and as resources allow.

\begin{figure}
    \centering
    \includegraphics[width=\linewidth]{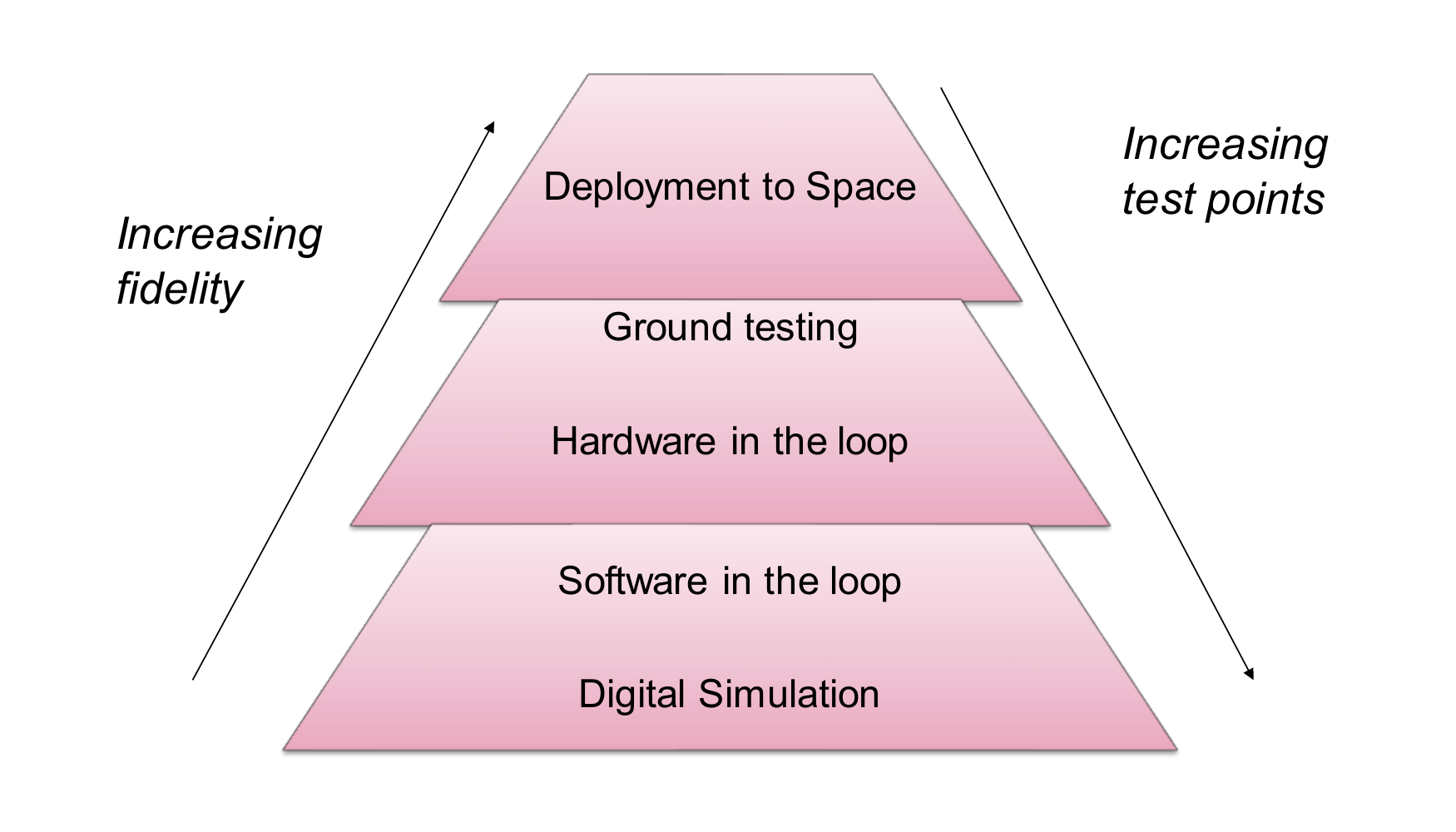}
    \caption{Number of test points exchanged for increased fidelity as testing moves up the system composition hierarchy.}
    \label{fig:testpoints}
\end{figure}

As shown in Fig.~\ref{fig:testpoints}, the framework makes a trade-off between the number of test points used and the fidelity of measuring the system. Moving up the hierarchy towards a completely realized system, testing achieves better fidelity, but becomes more difficult, and fewer test points are possible. 

\section{Test Design Framework}\label{framework}
The framework does not specify a series of tests to run. Instead, the framework helps inform comprehensive test plan design by outlining the considerations to address. 
These considerations will change at each slice in the VTP model leading to different test designs. Additionally, the ``P'' phase of the VTP model requires that information learned during operational test slices is incorporated into future tests plans.  For example, a system could employ a combinatorial interaction testing strategy with automated tests of each agent in its fielded state in an attempt to monitor if any had been compromised by an adversary.  The outcomes of these tests will be incorporated into future algorithm updates and result in a new round of independent tests to verify the successful implementation of updates to the agents. 

\begin{figure}
    \centering
    \includegraphics[width=.7\linewidth]{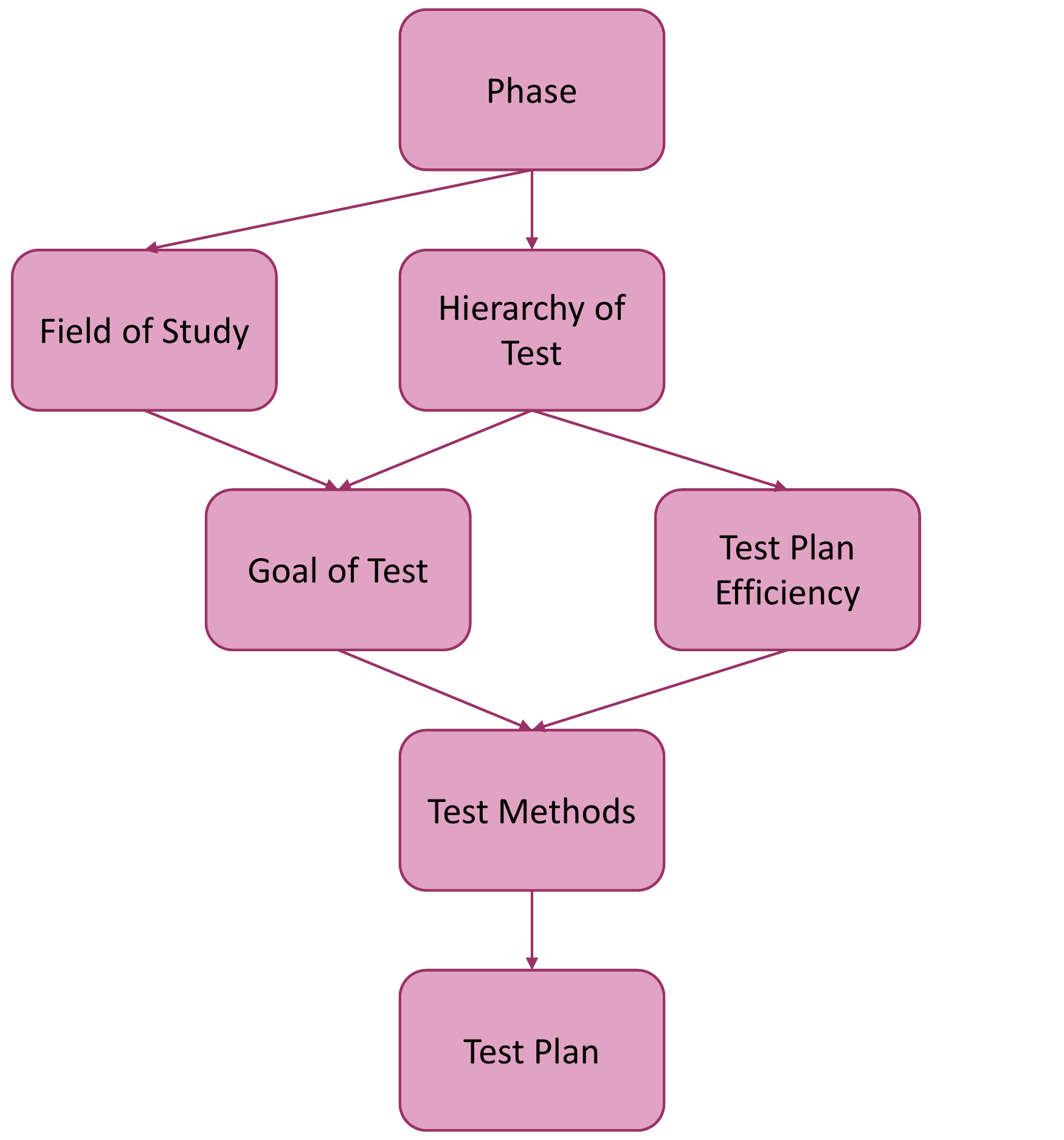}
    \caption{Process by which framework guides considerations towards a comprehensive test plan}
    \label{fig:frameworkprocess}
\end{figure}

The process by which the framework guides considerations to result in a test plan is graphically represented in Fig.~\ref{fig:frameworkprocess}. It begins with identification of the current phase of the life cycle for the system under test (SUT) and guides identification of the field of study and hierarchy of test. Hierarchy of test informs how to identify the components inside the SUT and which component levels should be tested. Identification of components, levels, and their interactions defines the test input space. Test plan efficiency guides the focus of the test given the cost of each test run at the current scale and assumes systems lower in the hierarchy function as expected. The framework process also includes identification of goals of the test and reasonable test methods. The field of study for the SUT and hierarchy of test inform the goal of the test. Goals are combined with test plan efficiency to determine appropriate test methods. Finally, the collected knowledge through all prior steps guides creation of the test plan.

\begin{figure*}
    \centering
    \includegraphics[width=\linewidth]{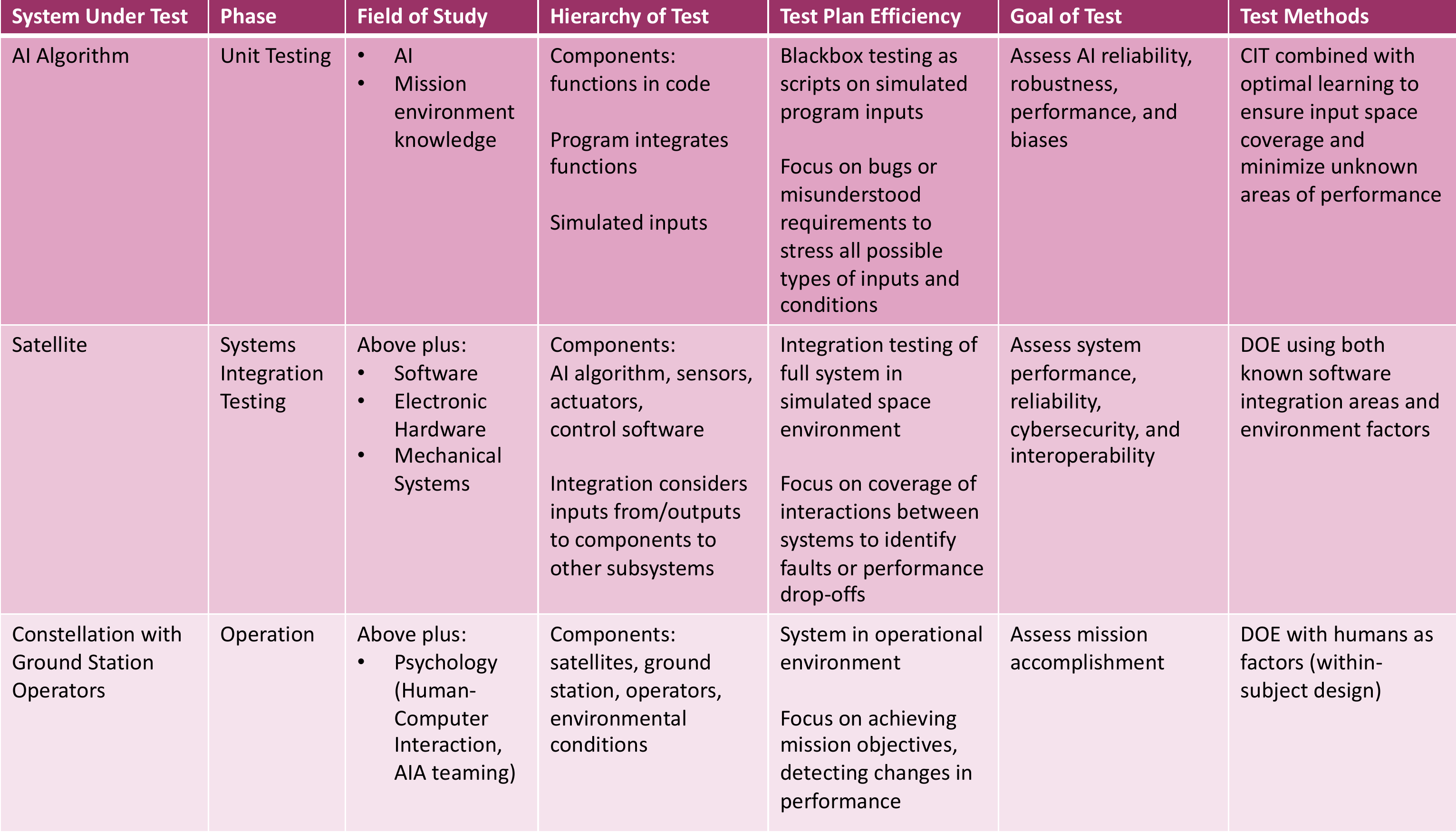}
    \caption{High-level overview of how each framework concept contributes to guiding test plan development}
    \label{fig:frameworkexample}
\end{figure*}

In Fig.~\ref{fig:frameworkexample}, three example scenarios within the multi-agent system of AIAs life cycle are provided. A non-exhaustive list of fields of study involved in the SUT and from which testing strategies should be drawn is given.  

Recall our use case, a satellite network tasked with conducting broad area search and point detection.  Fig.~\ref{fig:usecase} depicts the satellite network designed to observe shipping traffic.  The network conducts broad area search to understand normal traffic patterns and detect activities that could indicate potential illegal shipping activities.  Once an anomaly is detected, the network has multiple objectives: continue broad area search and maintain a track on the anomalous vessel.  Using this context, we can walk through Fig.~\ref{fig:frameworkexample} and propose reasonable test design structures. 

The first row identifies an early phase of testing focused on the AI algorithm.  This algorithm could be black-box or fully specified.  In our use case, consider the central controller, whose objective is to task the satellites in the network to both provide wide area coverage and maintain tracks.  Our test goal is to assess the reliability, robustness, performance, and biases of the central controller.  Because testing will be conducted via simulations, tests are relatively affordable, and we can afford a strategy that allows both comprehensive coverage via CIT and investigation of any areas of high uncertainty via optimal learning (tests can be sequentially added at low cost).  The CIT design could leverage historical shipping data crossed with CIT at a high strength (covering many interactions) to embed anomalous traffic into the historical data. The factors in the CIT test set may include vessel country of origin, vessel type, vessel size, and geospatial location.  Optimal learning is used to augment the CIT in areas of high uncertainty or that show large performance changes in the AI algorithm.

The second row of Fig.~\ref{fig:frameworkexample} shows how the considerations change when moving up to the satellite scale of testing. Testing must now consider not only algorithm performance, but also how that algorithm integrates with the system's additional software, electronic hardware, and mechanical systems. This necessitates identification of new factors and test design strategy. The knowledge gained in testing the AI via CIT/Optimal learning is leveraged to identify a subset of scenarios for input into the system-centric test design.  Here we may use a hardware-in-the-loop test facility where satellite sensors are given simulated inputs, but all additional aspects are real (e.g., simulating the space deployment).  Factors include AI performing scenario (potentially binned into low/medium/high based on the outcomes of the previous AI testing), active adversary (yes/no), and weather impact on inputs (e.g., clouds, rain), etc.  Experimental designs are used that focus on understanding how system performance changes as a function of the simulated inputs combined with various executions of the AI algorithm. 

Finally, once the system is deployed, we may need to understand mission accomplishment of the fully connected system.  In Fig.~\ref{fig:frameworkexample}, we focus on the human integration at a ground control station as important to mission accomplishment.  Here our test size is limited by the number of humans that are qualified mission controllers for the constellation.  We use a within-subject design to assign different ground teams to various scenarios on the deployed system.  The tasks are controlled by focusing the constellation on certain parts of the ocean for an operating period.  

The three scenarios provide a hypothetical example of how to use the elements in the test design framework, coupled with the slice in the VTP model to develop a test design strategy that pairs the goal of the test, derived from the field of study and the hierarchy of test with desired test efficiencies, for a complex system-of-systems at multiple scales.

\section{Conclusions and Future directions}\label{conclusions}
The framework we propose in this work specifies the three concepts of field of study, hierarchy of test, and test plan efficiency to be considered in each of the design, development, and deployment phases of the VTP model in order to guide the creation of a comprehensive test plan. 

This unifying framework synthesizes T\&E methodology and is generalizable to many contexts involving multi-agent systems of AIAs. We have provided examples throughout of how the framework would be applied to the use case of a constellation of satellites conducting both broad area search and point detection in a series of tests. Evaluation of the framework against additional use cases is needed to assess its usefulness for this category of complex systems and to identify any features needed to assist in guiding the creation of test plans. 

 Context-specific challenges in this work led to the formulation of two new research questions. Testing the integration of a system-of-systems composed of subsystems of the same type may lead to tests that appear ``symmetrical'' such that multiple tests effectively represent the same configuration. For example, consider testing the integration of a constellation of satellites having two satellites, $S_1$ and $S_2$, with identical settings for all components, but appearing in different locations in orbit, say $p_x$ and $p_y$. That is, they have the same payload, algorithms, control software, and actuators. If the satellites are listed as factors in a CIT test suite, without using constraints, tests should be generated including both combinations $\{(S_1,p_x),(S_2,p_y)\}$, $\{(S_1,p_y),(S_2,p_x)\}$ in larger strength interactions, though at early phases of testing before learning or damage has occurred, both satellites are interchangeable and thus both observations are not necessary. When testing is expensive, it may be desired to observe and analyze only the test cases needed. Thus, a test suite generation tool that avoids producing these kinds of symmetrical tests is needed. Using constraints to solve this challenge and generate a test suite can be computationally expensive in the presence of a large number of constraints such as for a large constellation. Taking inspiration from sequence covering arrays, we hypothesize that a partial ordering on specified factors could be used to build covering arrays without the need to remove symmetrical tests.
 
 By evaluating the framework in the context of the satellite use case, we propose to examine how much of the variance in success and failure is captured by this framework and whether mission success at the top of the hierarchy can be predicted by the results of testing at the low, task-scale of the hierarchy. As tests at the global scale may be costlier and, in some cases, only possible after the constellation has been launched into space, preventing most changes to the system, early warnings predicting mission-scale success or failure provide immense value.


\bibliographystyle{IEEEtran}
\bibliography{ref}

\end{document}